# SOFIA/FIFI-LS Full-disk [C II] Mapping and CO-dark Molecular Gas across the Nearby Spiral Galaxy NGC 6946

F. Bigiel[1,2], I. de Looze[3,4], A. Krabbe[5], D. Cormier[2,6], A. T. Barnes[1], C. Fischer[5], A. D. Bolatto[7], A. Bryant[5], S. Colditz[5], N. Geis[8], R. Herrera-Camus[9], C. Iserlohe[5], R. Klein[10], A. K. Leroy[11], H. Linz[12], L. W. Looney[13], S. C. Madden[6], A. Poglitsch[8], J. Stutzki[14], and W. D. Vacca[10]
[1] Argelander-Institut für Astronomie, Universität Bonn, Auf dem Hügel 71, D-53121 Bonn, Germany; bigiel@astro.uni-bonn.de
[2] Institut für theoretische Astrophysik, Zentrum für Astronomie der Universität Heidelberg, Albert-Ueberle Str. 2, D-69120 Heidelberg, Germany
[3] Sterrenkundig Observatorium, Ghent University, Krijgslaan 281—S9, B-9000 Gent, Belgium
[4] Department of Physics and Astronomy, University College London, Gower Street, London WC1E 6BT, UK
[5] Deutsches SOFIA Institut, Universität Stuttgart, Pfaffenwaldring 29, D-70569 Stuttgart, Germany
[6] AIM, CEA, CNRS, Université Paris-Saclay, Université Paris Diderot, Sorbonne Paris Cité, F-91191, Gif-sur-Yvette, France
[7] Department of Astronomy and Laboratory for Millimeter-Wave Astronomy, University of Maryland, College Park, MD 20742, USA
[8] Max-Planck-Institut für extraterrestrische Physik, Gießenbachstrasse 1, D-85748 Garching, Germany
[9] Departamento de Astronomía, Universidad de Concepción, Barrio Universitario, Concepción, Chile
[10] SOFIA-USRA, NASA Ames Research Center, MS N232-12, Moffett Field, CA 94035-1000, USA
[11] Department of Astronomy, The Ohio State University, 140 West 18th Avenue, Columbus, OH 43210, USA
[12] Max Planck Institute for Astronomy, Königstuhl 17, D-69117, Heidelberg, Germany
[13] Department of Astronomy, University of Illinois, 1002 W Green Street, Urbana, IL 61801, USA
[14] I. Physikalisches Institut der Universität zu Köln, Zülpicher Straße 77, D-50937, Köln, Germany
Received 2020 March 24; revised 2020 September 3; accepted 2020 September 7; published 2020 October 29

## Abstract

We present SOFIA/FIFI-LS observations of the [C II] 158 $\mu$m cooling line across the nearby spiral galaxy NGC 6946. We combine these with UV, IR, CO, and H I data to compare [C II] emission to dust properties, star formation rate (SFR), $H_2$, and H I at 560 pc scales via stacking by environment (spiral arms, interarm, and center), radial profiles, and individual, beam-sized measurements. We attribute 73% of the [C II] luminosity to arms, and 19% and 8% to the center and interarm region, respectively. [C II]/TIR, [C II]/CO, and [C II]/PAH radial profiles are largely constant, but rise at large radii ($\gtrsim$ 8 kpc) and drop in the center ("[C II] deficit"). This increase at large radii and the observed decline with the 70 $\mu$m/100 $\mu$m dust color are likely driven by radiation field hardness. We find a near proportional [C II]–SFR scaling relation for beam-sized regions, though the exact scaling depends on methodology. [C II] also becomes increasingly luminous relative to CO at low SFR (interarm or large radii), likely indicating more efficient photodissociation of CO and emphasizing the importance of [C II] as an $H_2$ and SFR tracer in such regimes. Finally, based on the observed [C II] and CO radial profiles and different models, we find $\alpha_{CO}$ to increase with radius, in line with the observed metallicity gradient. The low $\alpha_{CO}$ (galaxy average $\lesssim 2\, M_\odot\, pc^{-2}\, (K\, km\, s^{-1})^{-1}$) and low [C II]/CO ratios ($\sim$400 on average) imply little CO-dark gas across NGC 6946, in contrast to estimates in the Milky Way.

*Unified Astronomy Thesaurus concepts:* Spiral galaxies (1560); Interstellar medium (847); Interstellar dust (836); Interstellar line emission (844); Galaxy structure (622)

## 1. Introduction

The observation of emission from singly ionized carbon ([C II]) plays a central role in the study of the star-forming interstellar medium (ISM) in galaxies. This is largely due to the role of [C II] as one of the brightest cooling lines across the galaxy population and its accessibility across virtually all redshifts. Therefore, observations of the [C II] line are an important tool to study gas morphology, mass, kinematics, dynamics (e.g., virial mass estimates), and velocity dispersions. In particular, extensive studies of its role as a star formation tracer in the Milky Way and nearby galaxies serve as important benchmarks for interpretation and calibration.

The high luminosity of the [C II] line, up to a few percent of the total IR luminosity of a galaxy, is due to mainly two factors. First, the carbon atom has a low ionization potential of 11.3 eV (lower than that of hydrogen or other metals like oxygen or nitrogen), so that singly ionized carbon exists in various phases of the ISM: neutral gas (the cold neutral medium, CNM, and photodissociation regions, PDRs), H II regions, and the warm ionized medium (WIM). While the phase breakdown has been, and continues to be, a topic of active investigation (e.g., Madden et al. 1997; Cormier et al. 2012, 2019; De Looze et al. 2014), recent, comprehensive observations in the Milky Way and nearby spiral galaxies, including NGC 6946, indicate that the neutral medium, and in particular the dense PDRs, is the dominant source of emission (Pineda et al. 2013; Abdullah et al. 2017; Croxall et al. 2017; Sutter et al. 2019).

Second, in addition to the widespread occurrence of [C II] emission, the fine-structure splitting of its ground state leads to a relatively low excitation energy ($E/k \sim 92$ K corresponding to $\lambda \sim 158\, \mu$m), which is easily excited via (predominantly neutral) collisions in the neutral (CNM, PDR) gas. This provides an efficient cooling mechanism, in particular in dense PDRs, where gas is heated by massive star formation via the photoelectric effect on small dust grains and polycyclic aromatic hydrocarbons (PAHs).

This link between massive star formation and fine-structure line cooling motivates the role of [C II] emission as a star formation rate (SFR) tracer (e.g., Stacey et al. 1991; De Looze et al. 2014; Herrera-Camus et al. 2015; Kapala et al. 2015; Sutter et al. 2019). Emission at near/mid-infrared wavelengths from small dust grains (as a proxy for the heating rate via photoelectrons), as well as in the far-infrared (a proxy for reprocessed emission from massive stars heating the gas) should both be expected to scale with the cooling rate and therefore the [C II] emissivity. This makes [C II] a viable alternative to commonly used SFR tracers, such as total infrared (TIR) emission. Extragalactic observations, however, reveal significant variations in the [C II]/TIR ratio among galaxies. In particular, in infrared luminous environments, [C II] emission appears less luminous relative to that expected from TIR emission, tracing the heating rate by massive star formation in the ISM ("cooling line deficit," e.g., Malhotra et al. 1997; Graciá-Carpio et al. 2011; Díaz-Santos et al. 2017; Smith et al. 2017; Pineda et al. 2018).

Our new SOFIA/FIFI-LS [C II] observations of the nearby ($D \sim 7.7$ Mpc, see references in Leroy et al. 2019), molecular gas rich ($M_{H2} \sim 10^{9.6} M_\odot$, Leroy et al. 2013), actively star-forming (SFR $\sim 6\, M_\odot$ yr$^{-1}$, Leroy et al. 2019) and relatively face-on ($i \approx 33°$) spiral galaxy NGC 6946 provides a testbed to address these topics across a full galaxy disk. Our high resolution (15″, corresponding to ~560 pc) sets these observations apart from prior low-resolution work in this galaxy (Madden et al. 1993; Contursi et al. 2002) or observations covering only part of the disk (e.g., from Herschel KINGFISH, de Blok et al. 2016; Smith et al. 2017) and is in line with a recent SOFIA study of M51 (Pineda et al. 2018). NGC 6946 has a moderate metallicity gradient of $\sim -0.3$ dex/$r_{25}$ (central oxygen abundance $12 + \log(O/H) = 9.13$ and 9.05 at $0.4 \times r_{25}$, whereas the galaxy average is ~8.99, Moustakas et al. 2010), using the calibration by Kobulnicky & Kewley (2004), where the optical radius $r_{25} = 5\farcm74$ (see Leroy et al. 2013). This makes examining radial trends of the [C II] emissivity in this galaxy particularly interesting. For comparison to model predictions of the CO-to-H$_2$ conversion factor $\alpha_{CO}$ in Section 3.5, we convert metallicities to the calibration by Pettini & Pagel (2004) following Kewley & Ellison (2008).

The paper is organized as follows: we describe the new FIFI-LS [C II] observations and the ancillary data, as well as conversion of observables to physical quantities, in Section 2. Section 3.1 describes results from spectral stacking across the disk of NGC 6946 to make sensitive, average measurements of various quantities in different dynamical environments (spiral arm, interarm, and central region). We analyze average radial trends in Section 3.2 and discuss the neutral gas heating efficiency and [C II]–SFR correlations for individual beam-sized regions in Sections 3.3 and 3.4. Radial profiles of $\alpha_{CO}$ and the associated contribution of CO-dark gas are discussed in Section 3.5. Throughout the paper, we assume a position angle of 243° and an inclination of 33° (Leroy et al. 2013).

## 2. Data

### 2.1. SOFIA Observations

Our [C II] data were taken with the "Far Infrared Field-Imaging Line Spectrometer" (FIFI-LS) instrument (Colditz et al. 2018; Fischer et al. 2018) aboard the Stratospheric Observatory For Infrared Astronomy (SOFIA, Young et al. 2012). The observations were obtained as part of FIFI-LS guaranteed time as well as the open time project 04_0139. The data were taken on eight SOFIA flights, over three flight series between 2015 March and 2016 March, and were observed in symmetric chop mode with a full throw of 8′ and multiple chop angles to avoid chopping on emission. The coverage corresponds to a mosaic of 112 fields of $1′ \times 1′$. The galaxy was covered with 56 fields and then re-observed with an offset of 18″ and 6″ to create a half-pixel sampling as well as redundancy in the data set. The on-source integration time per pointing was 120 s, which resulted in 9.5 hr of total observing time including overheads. The point-spread function has a FWHM of ~15″ at 158 $\mu$m, and the spectral resolution is 270 km s$^{-1}$.

Data reduction was carried out using the FIFI-LS data reduction pipeline (Vacca et al. 2020), providing a data cube on a 3″ spatial grid. Flux calibration was performed from observations of calibration sources and the absolute amplitude calibration uncertainty was of the order 20%. The data were corrected for atmospheric absorption using transmission curves from ATRAN.[15] Each nod cycle was corrected with a transmission curve for flight altitude, telescope elevation, and a typical value for water vapor overburden (e.g., 7.3 $\mu$m in the zenith at 41,000 ft). The curve was convolved with the spectral resolution of the instrument ($R = 1200$). We trimmed the edge of the mosaic and removed noisy channels at the extremes of the spectral coverage. We then inspected the spectra across the map and identified the signal-containing channel range as $-400$ km s$^{-1}$ to 400 km s$^{-1}$. We removed a baseline by fitting a first-order polynomial to the remaining signal-free regions of the spectra on either side of this velocity range and subtracting the fit. The median $1\sigma_{rms}$ noise level per 0.0164 $\mu$m (~30 km s$^{-1}$) channel measured within the signal-free parts of the spectra is $8.4 \times 10^{-8}$ W m$^{-2}$ sr$^{-1}$ $\mu$m$^{-1}$.

To produce an integrated intensity map we followed a three step masking procedure. We first identified all voxels above a $5\sigma_{rms}$ level, and then grew this mask to include all channels down to $1\sigma_{rms}$ from these peaks and for each line of sight, respectively. Lines of sight outside the mask were filled with a narrow range of channels around the line center. This procedure returned a mask containing a high dynamic range of emission, while including a noise floor where no emission was present. A similar method was implemented as part of the Herschel KINGFISH program data reduction (DR3).[16] The masked cube was then collapsed within the signal-containing velocity range to produce the integrated intensity map. The uncertainty, $\sigma_I$, on the integrated intensity, $I$, was calculated as $\sigma_I = \Delta\lambda\, \sigma_{rms}\, N_{chan}^{1/2}$, where $N_{chan}$ is number of channels contributing to each line of sight and $\Delta\lambda$ is the channel width. The median $\sigma_I$ across the mapped region was $3.2 \times 10^{-9}$ W m$^{-2}$ sr$^{-1}$.

The [C II] and [O I] lines were also observed with partial coverage as part of the Herschel KINGFISH program (Kennicutt et al. 2011; Croxall et al. 2012). We use the moment maps from the Data Product Delivery DR3, and convolved and regridded them to 15″. A comparison of the PACS and FIFI-LS [C II] intensities in the region covered by KINGFISH in our target galaxy shows that they agree well; the median ratio of PACS over FIFI-LS [C II] intensities is ∼1.0 and the sensitivities are formally slightly better for the KINGFISH PACS spectroscopy, though the comparison is somewhat dependent on matching methodologies (median $\sigma_I$(PACS) across their mapped region is ∼1.3 × 10$^{-9}$ W m$^{-2}$ sr$^{-1}$).

---

[15] https://atran.arc.nasa.gov/cgi-bin/atran/atran.cgi
[16] https://irsa.ipac.caltech.edu/data/Herschel/KINGFISH/docs/KINGFISH_DR3.pdf

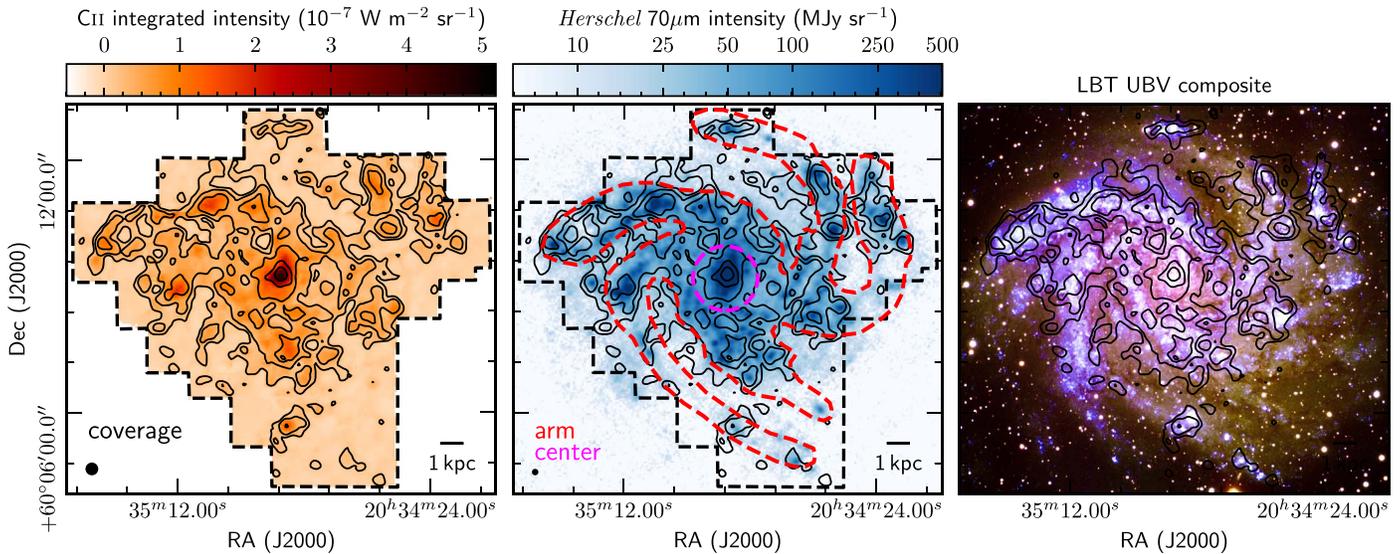

**Figure 1.** Left: SOFIA/FIFI-LS [C II] integrated intensity map with contour levels indicating signal-to-noise ratios of 4, 8, 16, 32, and 64 (corresponding to roughly 1.3, 2.6, 5.1, 10.2, and 20.4 $\times 10^{-8}$ W m$^{-2}$ sr$^{-1}$). These contours are overplotted in all panels. The beam size of 15″ is shown in the lower left, which is our working resolution. Middle: 70 $\mu$m emission from Herschel PACS on a logarithmic color scale and at native 6″ resolution. The extent of our SOFIA/FIFI-LS observations is shown by the black dashed contour. The three distinct regions referred to throughout are indicated by red and magenta contours (arms, center, respectively, and interarm). Right: LBT *UBV* composite map for comparison. The [C II] emission contours coincide tightly with the spiral structure in NGC 6946 (LBT image credit: D. Paris, V. Testa, LBC Team, and D. Thompson, LBTO).

### 2.2. Ancillary Data

We also use a variety of ancillary data sets, including 70 $\mu$m, 100 $\mu$m, and 160 $\mu$m Herschel KINGFISH photometry to compare the $I_{70}/I_{100}$ intensity ratio ("dust color") and the TIR intensity to our measurements. The latter is derived following Galametz et al. (2013). For compatibility with the recent literature, we estimate the star formation rate surface density using a hybrid tracer from a combination of GALEX far-UV and WISE band 4 22 $\mu$m emission following Leroy et al. (2019). In addition, as an alternative calibration, we use TIR emission following Murphy et al. (2011), calibrated specifically for NGC 6946 from 33 GHz free–free emission. We also make use of the Spitzer/IRAC data from SINGS (Kennicutt et al. 2003; Dale et al. 2007) to estimate the intensity of the emission from PAHs, which contribute to the gas heating. The PAH intensity, $I_{\rm PAH}$, is calculated using the IRAC 8.0 $\mu$m band, correcting for stellar emission with the IRAC 3.6 $\mu$m band as in Croxall et al. (2012), but not correcting for a warm dust contribution. Foreground stars and background galaxies have been removed using the mask from Muñoz-Mateos et al. (2009b) for this galaxy, which is based on IRAC colors and SExtractor identification. In addition, we compare to the PAH mass fraction "$q_{\rm PAH}$," derived from the dust modeling by Aniano et al. (2020) and based on the KINGFISH data products. We use the stellar mass map from M. Querejeta (2020, private communication) based on Spitzer IRAC 3.6 $\mu$m and 4.5 $\mu$m emission following Querejeta et al. (2015) and assume a mass-to-light ratio of 0.6 $M_\odot/L_\odot$ (Meidt et al. 2014). We apply the same mask to remove foreground stars.

CO intensities are derived from the HERACLES[17] $^{12}$CO(2−1) data (Leroy et al. 2009) and converted to $^{12}$CO(1−0) line intensities assuming a constant (2−1)/(1−0) line ratio of 0.7, appropriate for this galaxy (den Brok et al. 2020). We derive H$_2$ mass surface densities assuming a constant CO-to-H$_2$ conversion factor of 2.0 $M_\odot$ pc$^{-2}$(K km s$^{-1}$)$^{-1}$, about half the canonical Galactic value (e.g., Dame et al. 2001). We refer to Sandstrom et al. (2013) for a detailed study of the conversion factor including NGC 6946 specifically. We also use the THINGS[18] data products to compare to the distribution of atomic hydrogen (Walter et al. 2008; we also follow this paper regarding the derivation of mass surface densities). None of the quoted gas mass surface densities include the contribution from helium; however, all are deprojected assuming the inclination in Section 1.

All integrated intensity and mass surface density maps and data cubes are convolved to a common resolution of 15″ (corresponding to 560 pc, the $q_{\rm PAH}$ map is kept at its native 18″ resolution), put on a 15″ and 3″ pixel grid, aligned to the FIFI-LS [C II] astrometric grid and masked outside the field covered by our FIFI-LS mosaic (see Figure 1). We use the 15″ pixel grid data cubes and maps for our pixel plots, where individual data points thus represent largely independent measurements. For the stacking analysis in Section 3.1 and to derive the radial profiles in Section 3.2, we use the 3″ pixel grid set of maps.

### 3. Results and Discussion

Figure 1 shows the FIFI-LS [C II], Herschel KINGFISH 70 $\mu$m dust continuum, and an optical *UBV* composite map from the LBT with [C II] contours from FIFI-LS. The [C II] line

---
[17] http://www.iram-institute.org/EN/content-page-242-7-158-240-242-0.html
[18] http://www.mpia.de/THINGS/Overview.html

**Table 1**
Dust Color and [C II] Line Ratios from Stacked Measurements within the Inner $r = 1.5$ kpc ("Center"), the Spiral Arms, and Interarm Region in NGC 6946

|  | Central 1.5 kpc | Spiral Arms | Interarm |
|---|---|---|---|
| [C II]/TIR [·$10^{-3}$] | 2.79 ± 0.04 | 5.65 ± 0.08 | 12.31 ± 0.47 |
| [C II]/CO [·$10^2$] | 1.45 ± 0.02 | 4.11 ± 0.06 | 5.29 ± 0.20 |
| $I_{70}/I_{100}$ | 0.98 ± 0.05 | 0.80 ± 0.08 | 0.69 ± 0.23 |
| [C II]/$I_{\rm PAH}$ [·$10^{-2}$] | 1.26 ± 0.02 | 2.00 ± 0.03 | 4.02 ± 0.15 |

**Note.** The ratios were computed from intensities measured in W m$^{-2}$ sr$^{-1}$.

is detected over a large part of the area covered by FIFI-LS and with high significance (signal-to-noise ratio > 4) over most of the disk. The distribution of [C II] emission follows closely that of the warm dust and traces tightly the spiral arms in NGC 6946.

### 3.1. Spectral Stacking—Arm, Interarm, and Central Line Ratios

We probe three key ratios across the disk of NGC 6946: the [C II]/TIR and [C II]/PAH intensity ratios, often used as proxies for the photoelectric heating efficiency if the cooling from [O I] can be neglected (Tielens & Hollenbach 1985), and the [C II]/CO intensity ratio probing the star formation activity normalized to the bulk molecular gas reservoir. While we study these trends azimuthally averaged in Section 3.2, we focus here on the contrast between spiral arms, the interarm region, and the CO-bright, central 45″ radius (~1.5 kpc) region.

Because [C II] emission is faint between the spiral arms for individual lines of sight (compare Figure 1), we stack [C II] spectra for three distinct regions to improve the signal-to-noise ratio and compare different line ratios. We define these regions by hand, based on the 70 μm dust, CO, and optical emission. Magenta and red contours in Figure 1 outline the center of the galaxy and the spiral arms, respectively, and we refer to the remaining area as interarm in the following. We note that we have applied several alternative methods to define the regions for stacking, including contours based on TIR and CO intensity instead of selection by hand. These methods produce similar spectra and averaged ratios that are within the statistical uncertainties reported in Table 1. However, visual identification based on 70 μm, CO, and optical observations (Figure 1) lets us more reliably identify the spiral arms out to large radii.

We stack all [C II] and CO(2−1) spectra in each of these three regimes within the field covered by our [C II] observations (see Figure 1), i.e., we normalize all spectra according to the local mean velocity from a high signal-to-noise ratio tracer, here the CO(2−1), prior to averaging (see Jiménez-Donaire et al. 2017; Cormier et al. 2018, for details on the stacking procedure). To avoid averaging in noisy spectra, we require a peak above $3\sigma_{\rm rms}$ in at least one channel ±125 km s$^{-1}$ around the local mean velocity to consider a line of sight for the stack. Figure 2 shows the averaged [C II] spectra stacked in the three regimes, and Table 1 reports several relevant ratios of [C II], dust, and CO emission. The IR intensities are directly averaged in the respective integrated intensity maps for each region. The reported statistical errors are derived from propagating observational uncertainty (rms of the noise). For the [C II] ratios, we treat the [C II] error as the dominant source of uncertainty. For the dust color, we propagate statistical

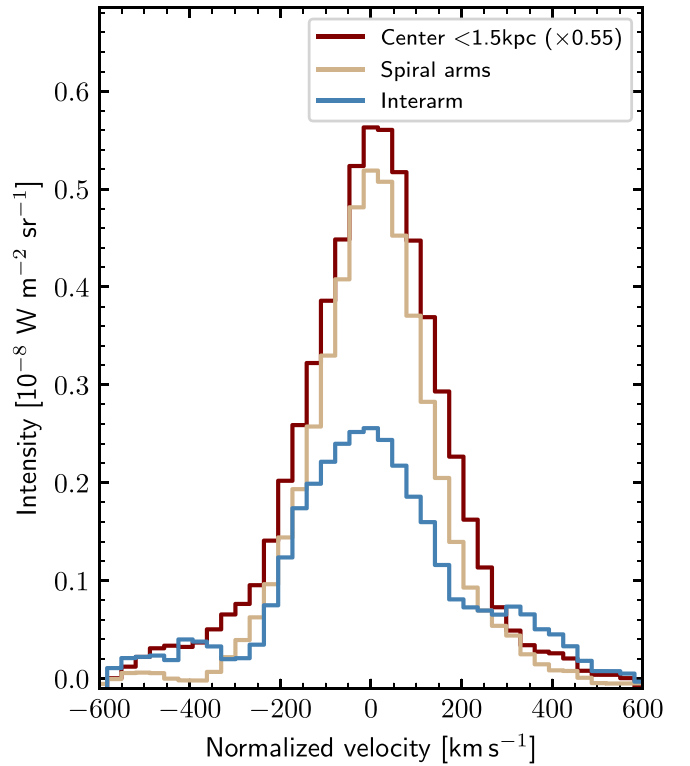

**Figure 2.** Stacked (velocity normalized and averaged) and scaled mean [C II] intensity spectra in NGC 6946 for the central $r = 1.5$ kpc region (red), the spiral arms (yellow), and the interarm region (blue). The scaling factor for the center spectrum is shown in the legend. Only lines of sight with at least one $3\sigma$ peak are included. Stacking allows to compare the average line shapes at high significance in these different environments and reveals different line widths, the largest of which is measured for the interarm regime, which contributes only ~8% to the overall [C II] luminosity of the galaxy.

uncertainty of the individual bands from the provided error maps.

The stacked spectra in Figure 2 show that the line widths differ depending on environment (note that the center spectrum is scaled): the interarm spectrum shows the largest line width (~360 km s$^{-1}$), followed by the center (~320 km s$^{-1}$) and the arm spectrum (~280 km s$^{-1}$). On the 560 pc scales we probe, line widths likely reflect larger scale gas dynamics like cloud-to-cloud dispersion and rotation, which drives the line width most severely in the center and the interarm regime. We note that the "bump" in the interarm spectrum at around 400 km s$^{-1}$ cannot be clearly associated with a galaxy feature and likely reflects a noisier region near the map edge.

While individual lines of sight are often detected at moderate signal-to-noise ratios, in particular in between the spiral arms, the stacking method yields a high-significance detection ($I/\sigma_I \simeq 25$) in this regime. This allows us to compare the mean interarm velocity-integrated intensity to that in the spiral arms; the former is about a factor of ~1.5 lower (the peak intensity is a factor of ~2 lower, see Figure 2). We also compute the line luminosity for the entire galaxy and in each region by stacking all spectra, i.e., without requiring a $3\sigma$ peak. The integrated [C II] galaxy luminosity is ~$7.7 \times 10^{41}$ erg s$^{-1}$. We find that 73% of the [C II] luminosity arises from the spiral arms, 19% from the central region, and 8% from the interarm region. This is compatible with Contursi et al. (2002), who attribute less than 40% of the [C II] emission in NGC 6946 (using ISO data) to diffuse gas, though we note that our definition of diffuse gas is somewhat different and based on morphological identification of interarm emission. In M51, Pineda et al. (2018) report a similar fraction, with 75% of the [C II] emission coming from the disk (arm and interarm).

Table 1 shows that the average [C II]/TIR, [C II]/CO, and [C II]/$I_{\rm PAH}$ ratios are different in the three regimes: they are lowest in the center, rise in the spiral arms, and are highest in between the spiral arms (though we note that the interarm ratios have the largest uncertainties). Typical [C II]/TIR ratios are ∼0.3–1.2%, which is a typical range for many nearby disk galaxies (Smith et al. 2017). [C II]/CO ratios are between 100 and 500. The [C II]/$I_{\rm PAH}$ ratio ranges between 1% and 4%. We reiterate that these stacked ratios are representative of lines of sight with at least a 3$\sigma$ peak in their spectrum and therefore, in particular in the interarm region, not of the entire area. Nonetheless, the large averaged areas across the disk of NGC 6946 include emission from different galactic environments and galactocentric radii (see Figure 1). The full range of values for these quantities is therefore much larger; we discuss this in detail in the following sections of the paper.

The low [C II]/TIR ratios in the center, also known as the [C II] deficit, are extensively discussed in Smith et al. (2017). They find that the deficit correlates with higher metallicity and higher star formation rate surface density across nearby galaxies. To follow this up, we use the ancillary [O I] 63 $\mu$m line observations from KINGFISH/PACS, which trace warmer/denser gas than [C II] (Tielens & Hollenbach 1985). We derive the [O I]-to-[C II] ratio in the center of NGC 6946, which is higher than in the disk with values around 0.5–1 compared to ∼0.2. This may argue for a significant contribution of the [O I] line to the combined cooling line emission, potentially leaving only a marginal overall cooling line deficit. The $I_{70}/I_{100}$ ratio is also higher in the center than in the disk. Therefore these ratio trends corroborate the scenario of the central region typically having higher density and/or temperature, and therefore having conditions that are expected to give rise to relatively underluminous [C II] emission. In addition, recent spectrally resolved observations of Galactic and extragalactic star-forming regions point to potentially nonnegligible optical depth of the [C II] line in such environments (via foreground or self-absorption, e.g., Ossenkopf et al. 2013; Langer et al. 2016; Okada et al. 2019; Guevara et al. 2020), possibly contributing to the [C II] deficit.

Regarding the [C II]-to-PAH ratios, a trend of decreasing values with increasing surface brightness has been reported in previous studies (Croxall et al. 2012; Lebouteiller et al. 2012). Similar reasoning as for [C II]/TIR applies for the different values of [C II]/$I_{\rm PAH}$ in the central region and the spiral arms of NGC 6946. Variations in the latter ratio are not as pronounced, probably because the PAH emission also gets fainter in the galaxy center. We discuss this further from Section 3.3 onward.

### 3.2. Radial Trends

In this section we compare the radial distribution of H I, H$_2$, [C II], PAH, and TIR emission, as well as the SFR surface density from a combination of far-UV and 22 $\mu$m emission across NGC 6946. Figure 3 shows these radial profiles, constructed from the mean intensities in concentric, tilted rings using the disk orientation parameters in Section 1. The error bars show the uncertainties on the mean for each point from propagating the rms of the observational noise. We plot the TIR, [C II], and PAH profiles in surface brightness units (left axis, [C II] and PAH intensity are scaled to the innermost TIR profile point) and the H$_2$, H I, and SFR (the latter is scaled to the innermost H$_2$ profile point) profiles in mass and SFR surface density units, respectively (right axis, corrected for inclination). The bin size (i.e., the ring width) is 9″, corresponding to roughly half the resolution. For each profile, the points were shifted slightly along the $x$-axis to facilitate comparison. We limit the profiles to radii < 10 kpc where the map coverage is largely complete. For the beam-sized regions analyzed from Section 3.3 onward, we plot measurements across the full extent of the map (out to ∼20 kpc).

The H$_2$, TIR, PAH, [C II], and SFR profiles track each other closely, declining roughly exponentially across most of the disk and showing a sharp rise toward the center. While the exponential decline is in line with Galactic observations, the latter reveals a central [C II] depression (Pineda et al. 2014), contrary to NGC 6946 and other nearby galaxies (e.g., de Blok et al. 2016). The H I profile is largely flat with radius, reflecting the limited dynamic range in $\Sigma_{\rm H\,I}$ across nearby galaxy disks (e.g., Bigiel et al. 2008, 2010; Walter et al. 2008). We quantify the radial decline of the former profiles by fitting exponentials to the radial profile points (simple least-squares minimization) and exclude the inner 5 profile points, where the profiles strongly deviate (indicated by the vertical black line at a galactocentric radius of 1.5 kpc). We find scale lengths of 4.5, 3.2, 4.0, 5.6, and 4.8 kpc for TIR, H$_2$, PAH, SFR, and [C II], respectively, reflecting a somewhat tighter correspondence between the radial TIR, PAH, and [C II] distributions, a slightly more severe decline of the H$_2$, and a slightly less severe decline of the SFR profiles across NGC 6946. This also implies varying [C II]/TIR, [C II]/H$_2$ (or [C II]/CO), [C II]/PAH, and [C II]/SFR ratios with radius, as shown in the right panel and in Figure 4 (and further discussed below and in the following sections). We note that a low, central CO-to-H$_2$ conversion factor (Sandstrom et al. 2013) would qualitatively lower, though not remove, the central H$_2$ excess.

de Blok et al. (2016) compared CO, H I, and [C II] profiles for 10 nearby galaxies, including NGC 6946. However, these were based on the limited disk coverage (typically major axis strips) of the Herschel KINGFISH survey (Kennicutt et al. 2011). For NGC 6946, this includes a major axis strip and several extranuclear pointings. Our full-disk profiles appear smoother in terms of radial structure, but are in excellent agreement in terms of measured surface brightnesses.

We measure the scatter in each radial bin, which is a combination of azimuthal (arm/interarm) and random scatter (noise). As expected, the scatter rises toward the outer part of the disk for all profiles except H I. The radial trend of the relative scatter about the mean [C II], TIR, PAH intensities and H$_2$ and SFR surface densities are quite comparable: the relative 1$\sigma$ rms scatter rises from about 10–20% in the inner part to ∼150–200% (a factor of ∼1.5–2) in the outer disk (H I has a roughly constant scatter of ∼40% across the disk).

As a coarse estimate of the astrophysical variation in [C II] intensity in each radial ring, we subtract the statistical rms scatter of the noise from the observed scatter (standard deviation) in quadrature. We find the estimated astrophysical variation exceeding the statistical noise by a factor of about 3–10, with a decreasing trend with radius. This means that there is a significant degree of astrophysical variation at each radius, which is particularly pronounced across the inner disk, and likely largely driven by azimuthal variations across arm-interarm environments (see the discussion in the following sections).

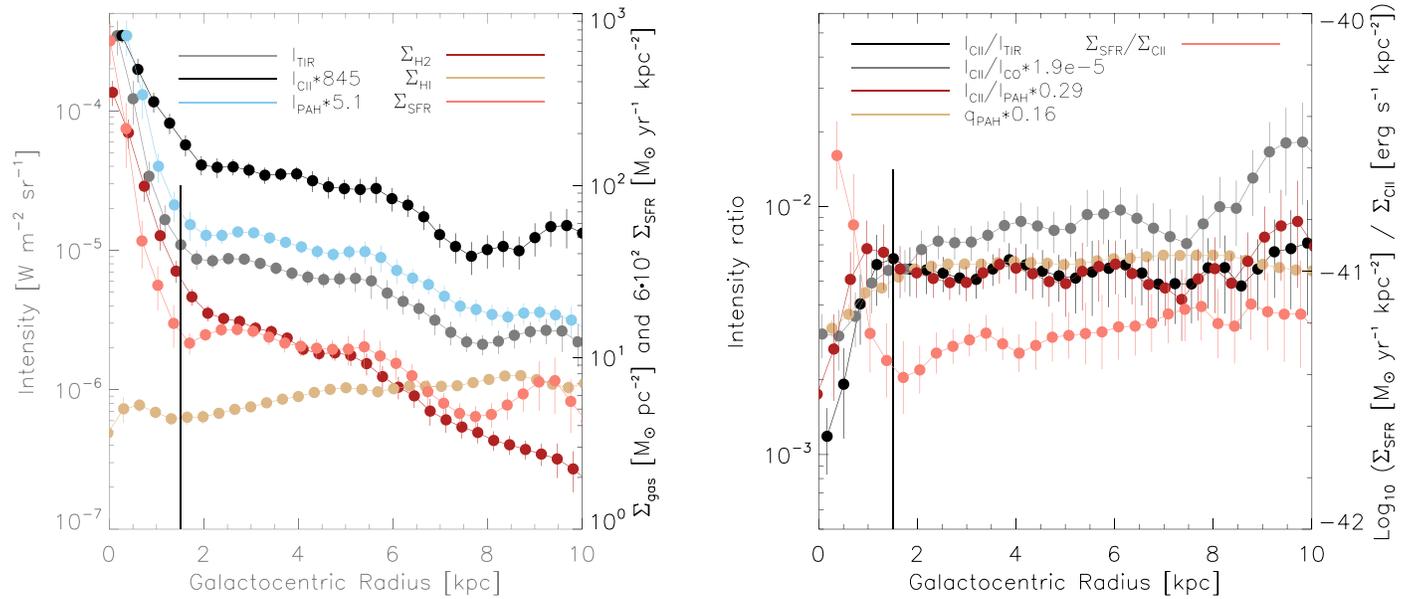

**Figure 3.** Left panel shows radial profiles of TIR, PAH, and [C II] intensities (left axis), as well as H I, H$_2$ (assuming a constant, roughly half Galactic CO-to-H$_2$ conversion factor), and SFR surface densities (right axis). The right panel shows several [C II]-intensity ratio profiles and the PAH mass fraction $q_{\rm PAH}$ (left axis), as well as the $\Sigma_{\rm SFR}/\Sigma_{\rm C\,II}$ ratio (right axis). The points in both panels are running means in steps of about one half-beamwidth (9″); the error bars show the (statistical) 1$\sigma$ uncertainty on the mean in each projected ring. For ease of comparison, the [C II], PAH, and TIR profiles are locked at the innermost point (left panel), and all ratio profiles except $\Sigma_{\rm SFR}/\Sigma_{\rm C\,II}$ are locked at the vertical line at $r = 1.5$ kpc (right panel, scaling factors of the profiles are provided in the legend). This line separates the two regimes of sharply rising or declining intensity and ratio profiles at small radii in NGC 6946 from the more smooth, exponential behavior across most of the disk.

The intensity and ratio profiles in both panels of Figure 3 show two distinct regimes separated by the black vertical line. There is an inner part extending to ∼1.5 kpc, where the TIR, [C II], and PAH intensities as well as the H$_2$ and SFR surface density increase strongly toward the center, whereas the ratios of [C II]/TIR, [C II]/PAH, and [C II]/CO in the right panel drop. For the large part of the disk beyond this radius, the intensities decline (and the ratios stay relatively constant or mildly rise) smoothly and exponentially; individual bumps, in particular in the SFR and [C II] profiles, correspond to prominent star-forming regions in the outer disk. For better visibility, we have locked all three ratio profiles in the right panel at that radius. In the outer disk, from about $r = 8$ kpc outward, all of the [C II] ratio profiles begin to rise. In the center, we see that the quantity that drops most toward small radii is [C II]/TIR, followed by [C II]/PAH. The PAH fraction, $q_{\rm PAH}$, decreases toward the center, where PAHs are likely destroyed by the more intense radiation field in the center, as observed previously (e.g., Muñoz-Mateos et al. 2009a; Aniano et al. 2012). The radial variation of $I_{\rm PAH}$ in the left panel clearly differs from the relatively flat radial trend observed for $q_{\rm PAH}$, and demonstrates that IRAC 8 $\mu$m emission does not necessarily trace the PAH mass fraction well. It is likely that hot dust continuum emission contributes significantly to the IRAC 8 $\mu$m band and/or that IRAC 8 $\mu$m emission is mostly sensitive to the UV radiation field responsible for the excitation of PAH molecules, rather than to the PAH dust mass fraction. Given the relatively constant PAH dust mass fraction outside the galaxy center, it is not surprising that the [C II]/TIR and [C II]/PAH intensity ratios follow each other closely across the entire disk. At larger radii, the PAH mass fraction drops slightly. This could be related to lower metallicities at large radii and hence lower PAH abundance relative to dust (Aniano et al. 2020).

The increase of [C II]/TIR with radius across the inner part and toward large radii is observed in several nearby spiral galaxies (e.g., Kramer et al. 2013; Kapala et al. 2015; Smith et al. 2017). The $\Sigma_{\rm SFR}/\Sigma_{\rm C\,II}$ ratio rises steeply in the inner part, reflecting the well known [C II] deficit. Across most of the disk, this ratio rises only slightly toward larger radii. This implies a ratio at least broadly consistent with being constant and, following common practice in the literature, translates into a power-law slope near unity relating both quantities. We address this in more detail and for beam-sized regions in Section 3.4 and Figure 5.

In the following sections, we focus on individual, beam-sized measurements to study the diverse environments probed by the extended SOFIA [C II] map locally and discuss further the efficiency of the gas heating and [C II] as a star formation rate tracer as a function of location in the galaxy.

### 3.3. Neutral Gas Heating Efficiency

Energetic electrons released from dust grains, contributing to the TIR emission, and from PAHs via the photoelectric effect triggered by far-UV photons, are generally the main heating sources of the neutral gas. Figure 4 shows the ratio of $I_{\rm C\,II}/I_{\rm PAH}$ intensity as a function of both the $I_{70}/I_{100}$ and the [C II]/TIR intensity ratios. We note that for these and the following plots we show all 15″ pixels (corresponding to roughly independent measurements), including those at large radii of the respective maps and focus on significant measurements $\gtrsim 5\sigma$. The black points show a running median and the error bars show the standard deviation to illustrate the scatter.

The ratio of $I_{\rm C\,II}/I_{\rm PAH}$ as a function of the $I_{70}/I_{100}$ ratio shows large scatter. The scatter is much larger than typically observed in similar galaxies (e.g., Croxall et al. 2012). It is likely that at least some of the scatter comes from the larger spatial coverage probing a wider range of astrophysical environments. There is a modest trend of decreasing $I_{C\,II}/I_{PAH}$ ratio, with a Spearman's rank correlation coefficient of −0.3 and a trend of increasing $I_{C\,II}/I_{PAH}$ values with increasing distance from the galaxy center for large radii (orange and red points). Similar trends are observed in other nearby spirals, and analogous trends are seen in [C II]/TIR (e.g., Croxall et al. 2012; Kapala et al. 2015). Croxall et al. (2012) find that, in the two galaxies NGC 1097 and NGC 4559, $I_{C\,II}/I_{PAH}$ is flat up to a color ratio of $I_{70}/I_{100} \approx 0.95$ and decreases beyond. They suggest that the decrease is most likely due to intense radiation fields in the galaxy centers leading to grain charging. We barely reach infrared colors above $I_{70}/I_{100} = 0.95$ in NGC 6946, suggesting that similar effects are likely not at play over most of the galaxy disk. In the very center, however, such a decrease with high $I_{70}/I_{100}$ ratios are observed (dark blue points in Figure 4, left panel), in line with this scenario.

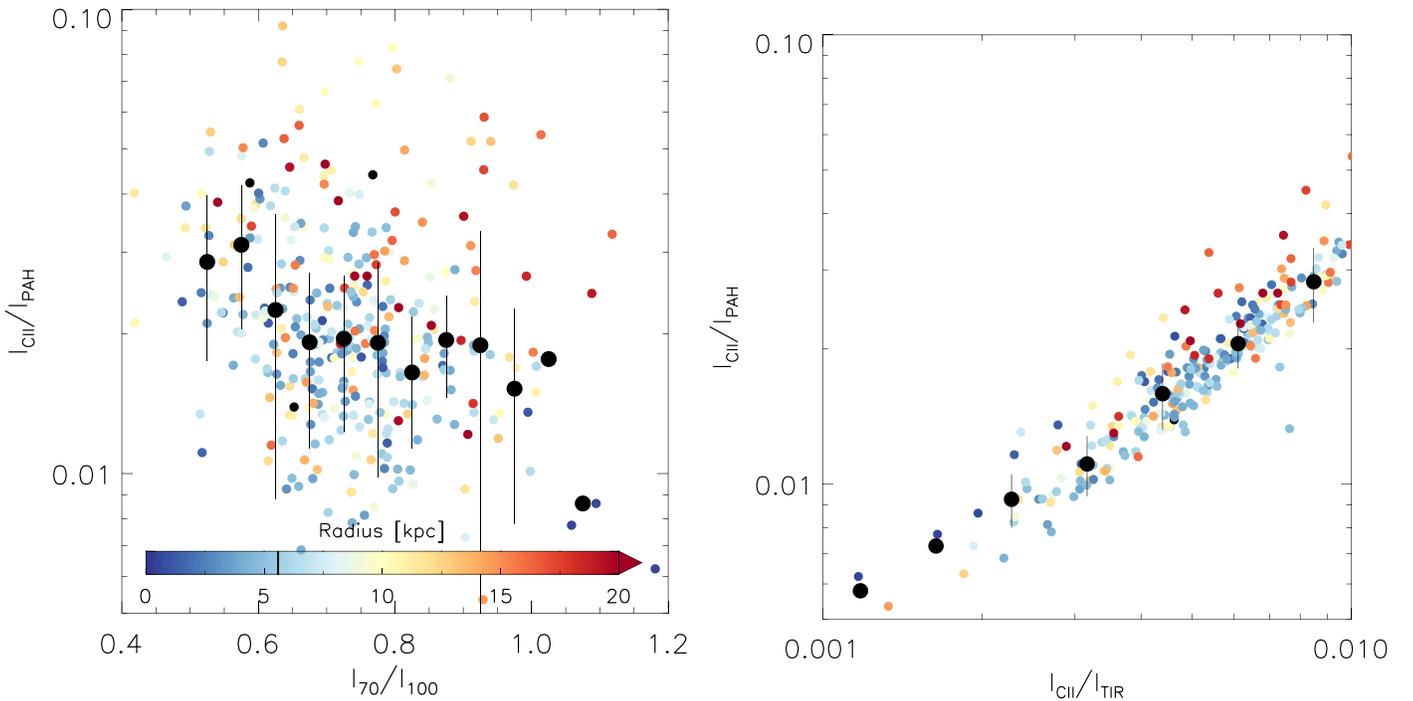

**Figure 4.** $I_{C\,II}/I_{PAH}$ intensity ratio as a function of $I_{70}/I_{100}$ (left) and $I_{C\,II}/I_{TIR}$ (right). Each point is a pixel in the 15″ maps (and therefore represents a roughly independent measurement) and color coded according to distance to the galaxy center. We focus on significant ($\gtrsim 5\sigma$) measurements only. We plot measurements across the full map out to radii of ~20 kpc and find elevated $I_{C\,II}/I_{PAH}$ ratios across the outer disk beyond ~10 kpc (and likewise, though not shown here, for the $I_{C\,II}/I_{TIR}$ ratio). The black points and error bars show a running median and standard deviation to illustrate the scatter.

Kapala et al. (2017) interpret the $I_{C\,II}/I_{TIR}$ variations in M31 (in particular at large radii) as a change in the relative hardness of the absorbed stellar radiation field (from different stellar populations, dust opacity, metallicity). The similar behavior in the radial trends for $I_{C\,II}/I_{PAH}$ (orange and red points in Figure 4, left panel, which are on average offset toward larger $I_{C\,II}/I_{PAH}$ ratios) suggests that variations in the relative hardness of absorbed stellar emission are likely also at play in NGC 6946. Given the tight correlation between $I_{C\,II}/I_{PAH}$ and $I_{C\,II}/I_{TIR}$ (right panel), a similar radial trend is found for data points beyond ~10 kpc for the $I_{C\,II}/I_{TIR}$ ratio (not shown). This tight correlation holds throughout the galaxy, with a Spearman's rank correlation coefficient of ~0.9 (Figure 4, right panel). There is more dispersion at high values, driven mainly by outer disk data points.

Overall, our data suggest that the PAH intensity tracks the [C II] intensity no better than the TIR; both ratios show comparable dynamic range and scatter (compare also Figure 3, right panel). Taken at face value, this may imply that PAHs are not more closely linked to neutral gas heating than larger dust grains in this galaxy. We note, however, that we have not taken into account the contribution from the ionized gas to the [C II] emission, which could contribute up to ~50% in metal-rich regions and may be almost negligible at low metallicity (Croxall et al. 2017; Cormier et al. 2019).

### 3.4. [C II] as a Star Formation Rate Tracer

Recent studies on large spatial scales or entire galaxies have shown that [C II] can be a useful tracer of the SFR in galaxies as it is one of the brightest cooling lines and accessible at many redshifts (e.g., De Looze et al. 2014; Capak et al. 2015; Herrera-Camus et al. 2015; Decarli et al. 2017). By probing the entire disk of NGC 6946, we investigate if [C II] tracks the SFR in a similar fashion in different environments across this galaxy.

Figure 5 shows the surface density of SFR from our UV-IR default SFR tracer, $\Sigma_{SFR}$, as a function of the [C II] surface density, $\Sigma_{C\,II}$, and the [C II]/CO intensity ratio for each beam-sized region in our map. Because of the large fraction of low signal-to-noise ratio measurements at low [C II] surface density, we use separate symbols in these plots for [C II] data points with signal-to-noise ratios greater than 3 (filled) and below that (open). To illustrate the overall trend, we overplot a running median and the $1\sigma$ rms scatter between $39.0 < \log \Sigma_{C\,II} < 40.5$, excluding the low intensity end where the distribution is dominated by low signal-to-noise ratio measurements (open circles).

Our data cover 2 dex of parameter space in the [C II]–SFR relation, with a clear transition from interarm to arm environment and most of the central lines of sight overlapping the luminous end of the arm distribution. The very central few lines of sight are clearly offset, showing a strong deficit of [C II] emission relative to their SFRs. The scatter in each bin (see error bars) is between 0.2 and 0.4 dex and is largest at higher luminosities.

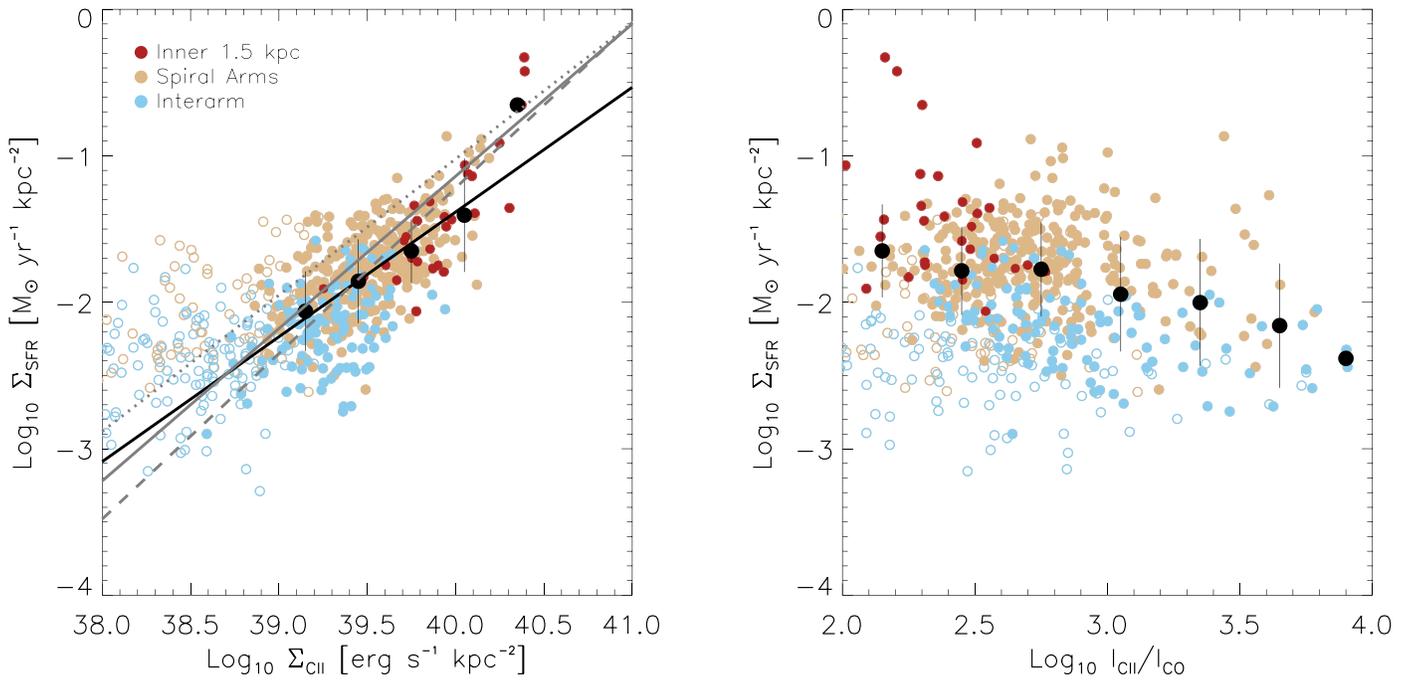

**Figure 5.** Star formation rate ($\Sigma_{SFR}$) as a function of [C II] surface density ($\Sigma_{C\,II}$, left) and [C II]/CO intensity ratio (right) for each 15″, roughly beam-sized measurement across the disk of NGC 6946. Data points are color coded by environment (see legend) and plotted as filled/open circles if their [C II] signal-to-noise ratio is greater/lower than 3. A fit to the filled data points is shown as a black line (details in text). Black circles and bars show a running median and scatter between $39.0 < \log \Sigma_{C\,II} < 40.5$. Gray lines show fits to resolved, kiloparsec-scale data from the Herschel KINGFISH and Beyond the Peak programs for disk galaxies (Herrera-Camus et al. 2015; Sutter et al. 2019, dashed and solid, respectively) and from the Herschel Dwarf Galaxy Survey (De Looze et al. 2014, dotted). The running median in the right panel is computed only for the significant measurements (filled circles).

The data distribution covers roughly the same range of [C II] and SFR surface densities than in other resolved, kiloparsec-scale studies of nearby galaxies (e.g., Herrera-Camus et al. 2015; Pineda et al. 2018). Because [C II]–SFR trends are commonly parameterized by power-law fits in the literature, we repeat this exercise for the purpose of comparing to prior work (though not for an exhaustive statistical characterization of these data). We carry out an ordinary least squares (OLS) fit treating [C II] surface density as the independent variable, assuming that the uncertainties associated with the measurement and calibration of $\Sigma_{SFR}$ are larger than those related to $\Sigma_{C\,II}$, and fit only significant measurements (filled circles). Both of these choices are motivated to be compatible to previous work in the field as much as possible. This fit in log–log space yields a slope of $0.85 \pm 0.05$ (statistical uncertainty) and an intercept of $-1.38$ measured at $\log \Sigma_{C\,II} = 40$. Excluding the center points (red circles) with often high SFR/[C II] ratios ("[C II] deficit") leads to a shallower slope for the disk points of $\sim 0.77$.

Repeating this exercise with our alternative SFR tracer (TIR, based on free–free emission calibrated for this galaxy specifically) yields a slope of $0.98 \pm 0.05$ and intercept $-1.23$. We conclude that systematics related to specific SFR calibrations leads to an uncertainty on the power-law slope at least by 0.15 dex, in line with the more comprehensive assessment in Herrera-Camus et al. (2015).

We note that the literature is diverse regarding statistical approaches characterizing such relationships. For instance De Looze et al. (2014) use a similar fitting method but a more rigorous signal-to-noise ratio cut, whereas Herrera-Camus et al. (2015) match the latter but fit an OLS bisector to their data. Hence we caution against overinterpreting a quantitative comparison. With this in mind, applying an OLS bisector fit to our distribution (for the default UV-IR SFR tracer) changes the slope by $\sim 0.36$ dex, increasing it to $\sim 1.21$. Using the same method, Herrera-Camus et al. (2015) derive aggregate slopes from bisector fits for 46 galaxies from KINGFISH (Kennicutt et al. 2011) between $\sim 1$ and 1.2, depending on the SFR tracer used. Their reference fit to the ensemble (with slope 1.13) is overplotted in Figure 5, left panel, as a gray dashed line and is in excellent agreement with this study. We derive a standard deviation about the fit of $\sim 0.25$ dex, which is only slightly larger than that ($\sim 0.21$ dex) measured by Herrera-Camus et al. (2015).

Sutter et al. (2019) fit measurements for 158 individual nuclear and extranuclear, kiloparsec-sized regions for 28 galaxies from the Herschel KINGFISH and Beyond the Peak (BtP) programs (Pellegrini et al. 2013). They derive SFRs from a combination of FUV and 24 $\mu$m emission and apply a Bayesian linear regression technique to derive a power-law slope of $1.04 \pm 0.05$ (solid gray line in Figure 5); we show their fit for all regions from the ionized and neutral medium. This appears consistent with our study, given the similar SFR tracer though different statistical treatment.

Comparing to measurements in the Milky Way, our fit also appears in line with the approximately linear fit derived from GOT-C+ Herschel [C II] data and 1.4 GHz free–free radio continuum data to trace the SFR across the Galactic plane (Pineda et al. 2014). Though we note that here, too, the fitting methodology differs. Finally, we compare our results to the [C II]–SFR relation of De Looze et al. (2014) for spatially resolved dwarf galaxies from the Herschel Dwarf Galaxy

Survey (gray dotted line in Figure 5). They estimate SFRs based on FUV and 24 μm emission similar to this study, and the range of [C II] surface densities covered by the resolved dwarf galaxy data is also quite comparable. Applying a more rigorous signal-to-noise ratio cut, they find an OLS fitted slope of 0.93, quite similar to this study of a gas-rich disk galaxy.

The [C II]/CO intensity ratio in the right panel of Figure 5 presents a broad range of values, between $\sim 10^2$ and $\sim 10^4$. These values are in the same range as those reported for Galactic regions (e.g., Pineda et al. 2013) and other metal-rich galaxies (e.g., Stacey et al. 1991, 2010). Data points belonging to the galaxy center largely overlap the measurements from the spiral arm, though the very central lines of sight are strongly offset toward higher SFRs. We also find a trend of decreasing $\Sigma_{SFR}$ with increasing [C II]/CO ratio, as illustrated by the overplotted running median (computed only including the significant measurements, i.e., the filled circles) and a correlation coefficient of $\sim -0.3$. Because the [C II]/CO ratio is particularly large where the SFR is low (and where the more diffuse, typically lower density gas in the interarm regions dominates, blue points), the relatively bright [C II] emission (compared to CO) makes it an important SFR and gas reservoir tracer in this regime, which contributes a nonnegligible fraction to the galaxy-integrated SFR (e.g., Foyle et al. 2010).

The large [C II]/CO ratios at large radii (compare also Figure 3, right panel) could at least be partly driven by a more efficient photodissociation of CO, caused by less efficient shielding. Interpreting the denominator in terms of $H_2$ surface density, a metallicity-dependent conversion factor increasing with radius (and thus at least on average with the [C II]/CO ratio) and likely also in the less enriched interarm regime, would qualitatively push in particular the blue points at the high [C II]/CO ratio end of the distribution toward lower ratios, possibly somewhat reducing the degree of correlation in this regime. A low central conversion factor in turn (Sandstrom et al. 2013) would move the offset, red points even further to the right, amplifying the disagreement with the disk measurements (see also the discussion in the next subsection).

### 3.5. The Spatially Resolved $\alpha_{CO}$ Conversion Factor and CO-dark Molecular Gas

CO emission is known to fail as an accurate tracer of the total molecular gas content under certain conditions. This applies specifically to low-metallicity regimes, where CO molecules are more easily photodissociated due to the reduced dust column and hard radiation, while $H_2$ is able to self-shield (e.g., Gnedin & Draine 2014). This "CO-dark" $H_2$ has been known to exist since the first FIR studies with the Kuiper Airborne Observatory (KAO) and Infrared Space Observatory (ISO) in low-metallicity dwarf galaxies, for which massive CO-dark, yet [C II]-bright, gas reservoirs were inferred (up to 10–100 times the mass of CO-bright molecular clouds, Poglitsch et al. 1995; Madden et al. 1997).

The detection of CO-dark gas in our Galaxy by comparing γ-ray observations (as a tracer of the total hydrogen gas, molecular and atomic) to CO and H I observations (Grenier et al. 2005; Ackermann et al. 2012; Hayashi et al. 2019) demonstrated that this is not only relevant for the low-metallicity interstellar medium. The presence of CO-dark gas in our Galaxy was later confirmed, e.g., from the Herschel GOT-C$^+$ observations (Langer et al. 2010, 2014; Pineda et al. 2013; Tang et al. 2016) and from infrared dust emission (Planck Collaboration et al. 2011; Reach et al. 2017).

The current paradigm assumes that CO-dark gas mostly consists of warm molecular gas, with a subdominant fraction from optically thick H I (e.g., Wolfire et al. 2010; Tang et al. 2016). Although dark in CO (and H I), these CO-dark gas reservoirs emit notably in [C II] and [C I] emission lines, which makes these lines potentially useful tracers of the CO-dark gas in galaxies (e.g., Papadopoulos & Greve 2004; Glover & Smith 2016; Bothwell et al. 2017; Valentino et al. 2020). In particular, the [C II]/CO ratio has been shown to be sensitive to the CO photodissociation efficiency in molecular clouds (e.g., Stacey et al. 1991; Wolfire et al. 2010; Accurso et al. 2017b; Madden et al. 2020).

Here, we make use of our full-disk [C II] and CO observations to estimate the variation of the $\alpha_{CO}$ conversion factor with galactocentric radius and constrain the CO-dark gas fraction across NGC 6946. The $\alpha_{CO}$ conversion factor is defined via

$$M_{H_2} [M_\odot] = \alpha_{CO} \times L_{CO} \text{ [K km s}^{-1}\text{ pc}^2], \quad (1)$$

with $\alpha_{CO}$ in units of $M_\odot$ pc$^{-2}$ (K km s$^{-1}$)$^{-1}$ (see Bolatto et al. 2013 for a review). In solar-metallicity environments, the canonical value is $\alpha_{CO} = 4.3 \, M_\odot$ pc$^{-2}$ (K km s$^{-1}$)$^{-1}$ (Bolatto et al. 2013), including the contribution from helium.

To derive a prescription for $\alpha_{CO}$, Accurso et al. (2017b) apply a Bayesian inference method to constrain dominant galaxy parameters driving variations in the observed [C II]/CO (1−0) ratios for galaxies covered by the xCOLD GASS (Saintonge et al. 2017) and Dwarf Galaxy Survey (DGS, Madden et al. 2013) programs. To that effect, the observational calibration between [C II]/CO(1−0), the metallicity 12+log(O/H), and the offset from the star-forming main sequence Δ(MS) is used in their model to derive a relation expressing $\alpha_{CO}$ as a function of 12+log(O/H) and Δ(MS). In combination with the multi-ISM phase radiative transfer simulations from Accurso et al. (2017a), they derive the following expression (Equation (24) from Accurso et al. 2017b):

$$\log(\alpha_{CO}) = 0.742 \log \frac{L_{[C\,II]}}{L_{CO(1-0)}}$$
$$- 0.944[12 + \log(O/H)]$$
$$- 0.109 \log \Delta(MS) + 6.439, \quad (2)$$

where Δ(MS) is (see their Equation (7))

$$\Delta(MS) = \frac{sSFR_{measured}}{sSFR_{ms}(z, M_\star)}. \quad (3)$$

The term sSFR is specific SFR, i.e., SFR/$M_\star$ and sSFR$_{ms}$(z, $M_\star$) corresponds to the analytical description of the main sequence from Whitaker et al. (2012):

$$\log(sSFR_{ms}(z, M_\star)) = -1.12 + 1.14z - 0.19z^2$$
$$- (0.3 + 0.13z) \times (\log M_\star - 10.5) \text{[Gyr}^{-1}]. \quad (4)$$

To estimate $\alpha_{CO}$ from these equations, we use the radial profiles observed for [C II], CO, and the SFR from Figure 3 and the radial metallicity gradient described in Section 1, following the Pettini & Pagel (2004) calibration. A radial profile for stellar mass surface density is constructed analogously to those in Figure 3, and we adopt a redshift of $z = 0$. The resulting

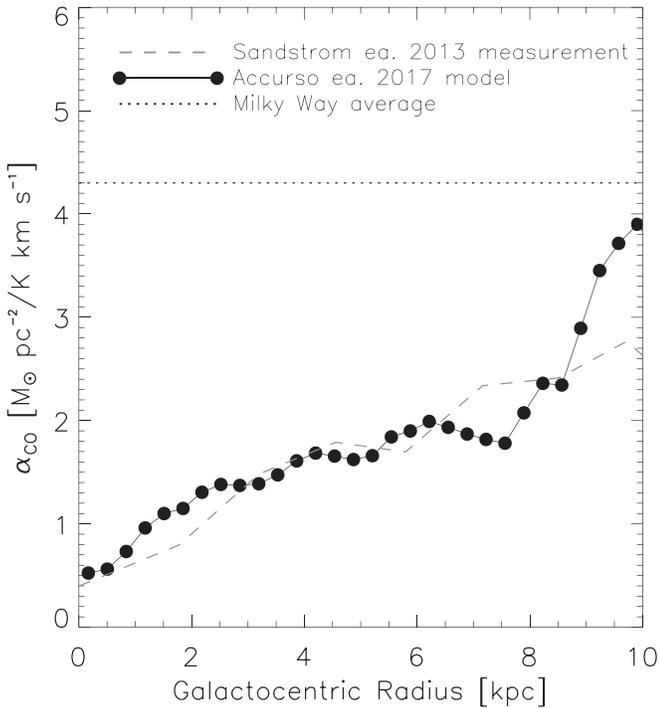

**Figure 6.** Radial profiles of $\alpha_{CO}$ for NGC 6946: canonical Milky Way average (dotted line), measurement from dust-to-gas ratios (Sandstrom et al. 2013, dashed), and prediction from the model by Accurso et al. (2017b, solid) using stellar mass (Section 2.2), our SFR, CO, and [C II] radial profiles from Figure 3, and the metallicity gradient (Section 1). The low values inferred from various models and observations (see main text) argue for a low CO-dark molecular gas fraction in this galaxy.

$\alpha_{CO}$ radial profile inferred for NGC 6946 is shown in Figure 6 as the black solid line.

While the Accurso et al. (2017b) recipes have been calibrated for whole galaxies (which we discuss further below), we note that their models are rooted in modeling individual star-forming regions. We thus consider applying their models to spatially resolved extragalactic observations at least an interesting exercise to provide further insight into how $\alpha_{CO}$ may vary across NGC 6946.

With these caveats in mind, we find the $\alpha_{CO}$ factor gradually increasing from low values of $\alpha_{CO} = 0.5\,M_\odot\,\mathrm{pc}^{-2}$ (K km s$^{-1}$)$^{-1}$ in the center to values reaching almost the Milky Way average of $\alpha_{CO} = 3.9\,M_\odot\,\mathrm{pc}^{-2}$ (K km s$^{-1}$)$^{-1}$ at galactocentric radii of 10 kpc according to the predictions from the Accurso et al. (2017b) model. The $\alpha_{CO}$ profile from Sandstrom et al. (2013) is overplotted as a dashed line for comparison. This profile is based on measurements of resolved dust-to-gas ratios and therefore is derived using an independent method. It is striking that there is an almost one-to-one correspondence between the dust-to-gas ratio based profile and the model trend (though the profile from Sandstrom et al. (2013) does not quite rise to similarly high values at large radii). In particular, the drop in $\alpha_{CO}$ in the central ~2 kpc of NGC 6946 is consistent between both approaches (also see Cormier et al. 2018 for updated radial $\alpha_{CO}$ profiles from Sandstrom et al. (2013) using measured, variable CO(2–1)/(1–0) ratios for some of their galaxies).

We note that estimating $\Delta$(MS) for the galaxy as a whole (and only using a radially varying [C II]/CO ratio and metallicity) moves the model curve down by roughly $\Delta\alpha_{CO} \sim -0.4$, except in the central ~1 kpc where it is identical. This leads to excellent agreement with the profile by Sandstrom et al. (2013) in the inner 2 kpc and beyond ~8 kpc, but puts the model curve slightly below their profile in between.

It is interesting that the low central values obtained from the model and the work by Sandstrom et al. (2013) are, however, a factor of a few below those inferred from virial mass studies of individual clouds in the center of NGC 6946 (Donovan Meyer et al. 2012). We refer to Sandstrom et al. (2013) for a more detailed discussion on the low, central $\alpha_{CO}$ values and diverging results from dust-based and virial mass estimates. Generally, they suggest the enhanced ambient ISM pressure affecting the virial balance, higher molecular gas temperature, and an enhanced diffuse gas component as possible explanations. For NGC 6946 specifically, dynamical broadening of the CO line reducing the optical depth in the center is a plausible explanation for the exceptionally low central $\alpha_{CO}$ value. In addition, recent work by Israel (2020) on CO, [C II], and [C I] emission in a large sample of nearby galaxy centers supports a scenario of low, central $\alpha_{CO}$ factors driven by elevated gas temperatures and large gas velocity dispersions (compared to the disk averages), but also attributes this at least in part to high central gas-phase carbon abundances.

Following Accurso et al. (2017b), comparing their models to integrated measurements across whole galaxies, we infer a global average $\alpha_{CO}$ factor based on the average metallicity 12 + log(O/H) = 8.66 (see Section 1, converted to the Pettini & Pagel 2004 calibration) and using the total star formation rate (log SFR = 0.79), stellar mass (log $M_\star$ = 10.5), and [C II]/CO ratio (396) from our measurements. The global conversion factor we derive from their models is $\alpha_{CO} = 1.4\,M_\odot\,\mathrm{pc}^{-2}$ (K km s$^{-1}$)$^{-1}$ and agrees well with the profiles in Figure 6 and the disk average of 2.0 from Sandstrom et al. (2013).

Another frequently used prescription is the metallicity-dependent $\alpha_{CO}$ relation from Amorín et al. (2016), which has a relatively weak metallicity dependence (slope $\beta \sim -1.45$). They infer the $\alpha_{CO}$ conversion factor from an empirical relation between the $H_2$ depletion timescale $\tau_{H_2}$ and the metallicity, where the $\tau_{H_2}$–SFR relation was derived for metal-rich galaxies in the COLD GASS survey (Saintonge et al. 2011), quite similar to NGC 6946. Their relation yields $\alpha_{CO} = 1.8\,M_\odot\,\mathrm{pc}^{-2}$ (K km s$^{-1}$)$^{-1}$, also consistent with the previous estimates. All these independent estimates have in common that the $\alpha_{CO}$ factor in NGC 6946 remains (at least) a factor of 2 below the canonical Galactic value of $\alpha_{CO} = 4.3\,M_\odot\,\mathrm{pc}^{-2}$ (K km s$^{-1}$)$^{-1}$, which also suggests that little CO-dark gas is present in this galaxy.

As a third model, we use the global [C II]/CO ratio from above (396) to constrain the CO-dark gas fraction using the models from Madden et al. (2020; in particular, see their Equations (3) and (4)). They infer CO-dark gas fractions using constraints from an extensive set of Cloudy models (Ferland et al. 2013) following Cormier et al. (2019). These models were calibrated on the observed [C II], [O I], CO(1–0), and TIR luminosities for DGS galaxies to infer a scaling relation for the CO-dark gas fraction as a function of the [C II]/CO ratio. We point out that the models are therefore primarily calibrated for less massive, lower-metallicity DGS galaxies with high [C II]/CO ratios ($\gtrsim$ 3000), which should be kept in mind for the following discussion. Qualitatively, a comparison with the CO-dark gas fraction in the Milky Way (see Madden et al. 2020, their Figure 9a) suggests that an extension of their relation to more massive galaxies will provide an upper limit on the CO-dark gas fraction, although the estimate from their models for NGC 6946 is already negligible (< 1%).

Generally, the low estimated CO-dark gas fractions from these models, as well as from the low $\alpha_{CO}$ estimates appear significantly lower compared to the fraction estimated in the Milky Way (20–70%, Langer et al. 2014) and in low-metallicity dwarf galaxies (> 75%, Poglitsch et al. 1995; Madden et al. 1997, 2020; Lebouteiller et al. 2019; Chevance et al. 2020). We speculate that most of the CO-dark gas, if any, is likely present in the outer regions and/or interarm regions of NGC 6946, where the [C II]/CO ratio is observed to be high with respect to the inner and/or arm regions of the galaxy (see Figures 3 and 5).

We note that we restrict our analysis in this section to galactocentric radii of < 10 kpc (or ~1.7 $r_{25}$) due to the incomplete coverage of the radial profiles further out (in particular for [C II], see Section 3.2). Even though we cannot rule out that a significant fraction of CO-dark gas is present at radii beyond 10 kpc, we consider a massive CO-dark gas reservoir in this galaxy unlikely, given that the molecular gas fraction drops significantly at large radii and given that the CO-dark gas fraction peaks between 3.5 kpc and 7.5 kpc in the Milky Way. The high metal abundance (and dust column) in NGC 6946 seems to prevent most of the CO molecules from being easily photodissociated. The absence of a massive CO-dark gas reservoir in NGC 6946 is in agreement with the lack of any evidence for a significant CO-dark gas reservoir in the metal-rich galaxy M31, as inferred from the dust-to-gas ratios in this galaxy (Smith et al. 2012).

## 4. Summary

We briefly summarize our results, comparing full-disk SOFIA/FIFI-LS [C II] mapping across the nearby spiral galaxy NGC 6946 to ancillary HERACLES CO and THINGS H I, as well as Spitzer and Herschel IR photometry, UV-IR and total IR-based SFR calibrations, and including the measured metallicity gradient and stellar mass distribution. We find that [C II] emission tracks the warm dust (70 $\mu$m) in the spiral arms closely across the disk. Stacking all [C II] spectra in the spiral arms, interarm regions, and the central ~1.5 kpc ("center"), respectively, allows us to make high significance, average measurements in these regimes. We attribute about 73% of the [C II] emission to the spiral arms and only 19% and 8% to the center and the interarm region. Focusing on lines of sight with significant [C II] detections only, the interarm stacked spectrum has the largest line width, whereas the spiral arm spectrum has the smallest (~360 km s$^{-1}$ versus ~280 km s$^{-1}$). [C II]/TIR, [C II]/CO, and [C II]/$I_{PAH}$ ratios change systematically: they are lowest in the center and highest in the interarm.

Studying radial trends of some of the key line ratios, we find relatively constant or slightly increasing [C II]/TIR, [C II]/CO, [C II]/PAH, and $\Sigma_{SFR}/\Sigma_{C\,II}$ ratios with radius over most of the disk. Excepting $\Sigma_{SFR}/\Sigma_{C\,II}$, they increase at larger radii (> 8 kpc) and drop in the center ($\Sigma_{SFR}/\Sigma_{C\,II}$ rises steeply). This "[C II] deficit" is in line with a higher density and warmer medium in the galaxy center. We estimate an "astrophysical" variation exceeding the statistical noise in each radius bin by factors of a few, arguing for significant local variation in each radius bin (likely driven largely by arm-interarm contrast).

Comparing these ratios for individual, kiloparsec-scale line-of-sight measurements across the galaxy disk, we find a trend of decreasing [C II]/PAH (and also [C II]/TIR) ratios with the 70 $\mu$m/100 $\mu$m dust color (particularly pronounced in the center). Measurements at largest radii ($r < 20$ kpc) are offset toward high [C II]/PAH ratios. Both trends could be related to the radiation field hardness, in line with previous work. TIR and PAH intensities track [C II] emission equally well with radius, which at face value might imply that PAHs and larger dust grains play a similar role in neutral gas heating in this galaxy.

Assessing the [C II]–SFR scaling relation, we reproduce the proportionality between both quantities commonly measured in the literature. Matching methodology as much as possible, we reproduce in particular the Herschel KINGFISH results by Herrera-Camus et al. (2015). We find that the choice of SFR tracer and statistical methodology (fitting method) has significant impact on derived statistical quantities, also in line with previous work. Splitting by environment, arm, interarm, and most of the central lines of sight appear to follow the same scaling relation, though the very center is offset toward high SFRs. We also show that [C II] becomes increasingly luminous relative to CO where the SFR is low (e.g., in the interarm regions) and that the three regions separate more clearly for these quantities. We consider more efficient photodissociation of CO at large radii and in between spiral arms as the most likely explanation. This emphasizes the importance of [C II] observations tracing molecular gas and SFR in such regimes.

We infer estimates for the $\alpha_{CO}$ conversion factor based on our observed [C II] and CO full-disk radial profiles, using the Accurso et al. (2017b) model and compare to prior observational work based on dust-to-gas ratio measurements (Sandstrom et al. 2013). We obtain an increasing trend for $\alpha_{CO}$ with galactocentric radius, in line with the metallicity gradient in this galaxy. The resolved $\alpha_{CO}$ factors are (at least) a factor of 2 below the Milky Way value (4.3 $M_\odot$ pc$^{-2}$ (K km s$^{-1}$)$^{-1}$), supported by equally low estimates for the global, average conversion factor from additional models and by the relatively low [C II]/CO ratios in this galaxy. We speculate that the high abundance of metals (and dust) prevents efficient photodissociation of CO molecules. These results imply a low, and potentially almost negligible, CO-dark gas fraction in NGC 6946, which is substantially different than in the Milky Way where the CO-dark gas fraction is quite significant.

Based on observations made with the NASA/DLR Stratospheric Observatory for Infrared Astronomy (SOFIA). SOFIA is jointly operated by the Universities Space Research Association, Inc. (USRA), under NASA contract NAS2-97001, and the Deutsches SOFIA Institut (DSI) under DLR contract 50 OK 0901 to the University of Stuttgart. We thank the staff of the SOFIA Science Center for their help. We thank G. Aniano, B. Draine, and collaborators for making available their $q_{PAH}$ map of NGC 6946 prior to publication, D. Paris, V. Testa of the LBC Team, and D. Thompson of LBTO for providing their optical composite map, J.C. Muñoz-Mateos for providing a copy of his source mask, M. Querejeta for sharing his stellar mass map ,and K. M. Sandstrom for providing their resolved $\alpha_{CO}$ profile. F.B. and A.T.B. acknowledge funding from the European Union's Horizon 2020 research and innovation program (grant agreement no. 726384/EMPIRE). I.D.L. gratefully acknowledges the support of the Research Foundation Flanders (FWO).

## ORCID iDs

F. Bigiel ● https://orcid.org/0000-0003-0166-9745
D. Cormier ● https://orcid.org/0000-0002-1046-2685
C. Fischer ● https://orcid.org/0000-0003-2649-3707
A. D. Bolatto ● https://orcid.org/0000-0002-5480-5686
C. Iserlohe ● https://orcid.org/0000-0003-4223-7439
A. K. Leroy ● https://orcid.org/0000-0002-2545-1700
L. W. Looney ● https://orcid.org/0000-0002-4540-6587
S. C. Madden ● https://orcid.org/0000-0003-3229-2899
A. Poglitsch ● https://orcid.org/0000-0002-6414-9408
W. D. Vacca ● https://orcid.org/0000-0002-9123-0068